\documentclass[12pt]{article}
\usepackage{amsfonts,amssymb,amsmath}
\usepackage[dvips]{graphicx}
\usepackage{color}
\newtheorem{theorem}{Theorem}

\begin{document}

\def\prg#1{\medskip\noindent{\bf #1}}  \def\ra{\rightarrow}
\def\lra{\leftrightarrow}              \def\Ra{\Rightarrow}
\def\nin{\noindent}                    \def\pd{\partial}
\def\dis{\displaystyle}                \def\inn{\,\rfloor\,}
\def\grl{{GR$_\Lambda$}}               \def\vsm{\vspace{-9pt}}
\def\Lra{{\Leftrightarrow}}
\def\cs{{\scriptstyle\rm CS}}          \def\ads3{{\rm AdS$_3$}}
\def\Leff{\hbox{$\mit\L_{\hspace{.6pt}\rm eff}\,$}}
\def\bull{\raise.25ex\hbox{\vrule height.8ex width.8ex}}
\def\ric{{(Ric)}}                      \def\tric{{(\widetilde{Ric})}}
\def\tmgl{\hbox{TMG$_\Lambda$}}
\def\Lie{{\cal L}\hspace{-.7em}\raise.25ex\hbox{--}\hspace{.2em}}
\def\sS{\hspace{2pt}S\hspace{-0.83em}\diagup}   \def\hd{{^\star}}
\def\dis{\displaystyle}

\def\hook{\hbox{\vrule height0pt width4pt depth0.3pt
\vrule height7pt width0.3pt depth0.3pt
\vrule height0pt width2pt depth0pt}\hspace{0.8pt}}
\def\semidirect{\;{\rlap{$\supset$}\times}\;}
\def\first{\rm (1ST)}
\def\second{\hspace{-1cm}\rm (2ND)}
\def\bm#1{\hbox{{\boldmath $#1$}}}
\def\nb#1{\marginpar{{\large\bf #1}}}

\def\G{\Gamma}        \def\S{\Sigma}        \def\L{{\mit\Lambda}}
\def\D{\Delta}        \def\Th{\Theta}
\def\a{\alpha}        \def\b{\beta}         \def\g{\gamma}
\def\d{\delta}        \def\m{\mu}           \def\n{\nu}
\def\th{\theta}       \def\k{\kappa}        \def\l{\lambda}
\def\vphi{\varphi}    \def\ve{\varepsilon}  \def\p{\pi}
\def\r{\rho}          \def\Om{\Omega}       \def\om{\omega}
\def\s{\sigma}        \def\t{\tau}          \def\eps{\epsilon}
\def\nab{\nabla}      \def\btz{{\rm BTZ}}   \def\heps{\hat\eps}

\def\tG{{\tilde G}}   \def\cF{{\cal F}}
\def\cL{{\cal L}}     \def\cM{{\cal M }}   \def\cE{{\cal E}}
\def\cH{{\cal H}}     \def\hcH{\hat{\cH}}
\def\cK{{\cal K}}     \def\hcK{\hat{\cK}}  \def\cT{{\cal T}}
\def\cO{{\cal O}}     \def\hcO{\hat{\cal O}} \def\cV{{\cal V}}
\def\tom{{\tilde\omega}}  \def\cE{{\cal E}}
\def\cR{{\cal R}}    \def\hR{{\hat R}{}}   \def\hL{{\hat\L}}
\def\tb{{\tilde b}}  \def\tA{{\tilde A}}   \def\tv{{\tilde v}}
\def\tT{{\tilde T}}  \def\tR{{\tilde R}}   \def\tcL{{\tilde\cL}}

\def\bA{{\bar A}}   \def\bB{{\bar B}}       \def\bC{{\bar C}}
\def\bE{{\bar E}}

\def\he{{\hat e}}   \def\hom{{\hat\omega}}
\def\hnab{{\hat\nabla}}     \def\hT{{\hat T}}

\def\nn{\nonumber}                    \def\vsm{\vspace{-8pt}}
\def\be{\begin{equation}}             \def\ee{\end{equation}}
\def\ba#1{\begin{array}{#1}}          \def\ea{\end{array}}
\def\bea{\begin{eqnarray} }           \def\eea{\end{eqnarray} }
\def\beann{\begin{eqnarray*} }        \def\eeann{\end{eqnarray*} }
\def\beal{\begin{eqalign}}            \def\eeal{\end{eqalign}}
\def\lab#1{\label{eq:#1}}             \def\eq#1{(\ref{eq:#1})}
\def\bsubeq{\begin{subequations}}     \def\esubeq{\end{subequations}}
\def\bitem{\begin{itemize}}           \def\eitem{\end{itemize}}
\renewcommand{\theequation}{\thesection.\arabic{equation}}

\title{Holography in Lovelock Chern-Simons AdS Gravity}

\author{Branislav Cvetkovi\'c{$^a$}, Olivera Miskovic{$^b$},  and   Dejan Simi\'c{$^a$}\footnote{
       Email addresses: {\tt olivera.miskovic@pucv.cl, cbranislav@ipb.ac.rs, dsimic@ipb.ac.rs}} \\
       {$^a$}Institute of Physics, University of Belgrade,\\
                      P. O. Box 57, 11001 Belgrade, Serbia\\
{$^b$}Instituto de F\'isica, Pontificia Universidad Cat\'olica de Valpara\'iso,\\
Casilla 4059, Valpara\'iso, Chile}

\date{\today}
\maketitle

\begin{abstract}
We analyse holographic field theory dual to Lovelock Chern-Simons AdS Gravity in higher dimensions using first order formalism.
We first find asymptotic symmetries in the AdS sector showing that they consist of local translations, local
Lorentz rotations, dilatations and non-Abelian gauge transformations. Then, we compute $1$-point functions of energy-momentum and spin currents
in a dual conformal field theory and write Ward identities. We find that the holographic
theory possesses Weyl anomaly and also breaks non-Abelian gauge symmetry at the quantum level.
\end{abstract}
\section{Introduction}

The AdS/CFT correspondence \cite{1} relates the fields in ($d+1$)-- dimensional asymptotically anti-de Sitter (AAdS) space
and correlators in a $d$-dimensional Conformal Field Theory (CFT). These two theories are dual in the asymptotic sector of gravity, such that the weak coupling regime of one is related to the strong coupling regime of another. For a weak gravitational coupling, the bulk theory is well-described by its semiclassical approximation, leading to the form of the duality most often used.

Since its discovery, the correspondence tools have been applied to many strongly coupled systems, giving rise to new insights into their dynamics, for example in hydrodynamics \cite{2} and condensed matter systems such as superconductors \cite{3}.

On the other hand, much effort has been invested in analysing the duality in semiclassical approximation of a bulk theory, with twofold purpose. First, it enables to test the conjecture itself. Second, it helps us to gain the knowledge about strongly coupled systems which are non-perturbative and not very well understood. However, most of this investigation deals with Riemannian geometry of bulk spacetime, see for example \cite{3,4,5,6,7,8}, while a more general structure based on Riemann-Cartan geometry, where both torsion and curvature determine gravitational dynamics, is mostly underinvestigated. One of the first studies of Riemann-Cartan holography used first order formalism to obtain a holographic dual of Chern-Simons AdS gravity in five dimensions \cite{9}. After that, in three dimensions, holographic dual to the Mielke-Baekler model was analysed in \cite{10}, and to the most general parity-preserving three-dimensional gravity with propagating torsion in \cite{11}. The physical interpretation of torsional degrees of freedom as carriers of a non-trivial gravitational magnetic field in 4D Einstein-Cartan gravity was discussed in
 \cite{12}.

Studying holographic duals of gravity with torsion has many benefits. Since its setup is more general, it also contains the results of torsion-free gravity. One of the very important features is that treating vielbein and spin connection as independent dynamical variables simplifies calculations
to great extent. In Ref. \cite{11}, it was shown that for three dimensional bulk gravity  conservation laws of the boundary theory take the same form in Riemann-Cartan and Riemannian theory when the boundary torsion is set to zero. Thus, it is possible to treat vielbein and spin connection as independent dynamical variables and reproduce Riemannian results in the limit of zero torsion. In this work, we extend the results of \cite{11} to all odd dimensions in case of holographic theory dual to Lovelock-Chern-Simons AdS gravity, by reproducing the conservation laws with respect to diffeomorphisms, Weyl and local Lorentz symmetry using first order formalism after taking a Riemannian limit.

Working in the framework of gravity with torsion also leads to richer boundary non-Abelian symmetries, as it is explicitly demonstrated for the particular model studied in this paper.

We  analyse a holographic structure of Lovelock Chern-Simons AdS Gravity \cite{13,14} in asymptotically AdS spaces. The key feature of this model is that it possesses a unique AdS vacuum, which is multiply degenerate in odd  $D\geq 5$ dimensions. Unlike general Lovelock-Lanczos \cite{15} gravity, it contains only two free parameters -- gravitational constant $\k$ and the AdS radius $\ell$.
This theory also features a symmetry enhancement from local Lorentz to AdS gauge symmetry.
Degenerate vacuum makes the linear perturbation analysis not applicable around the AdS background.
The holographic study in AAdS spacetimes, however, is non-perturbative, because  the gravitational fields in a dual theory are not dynamical but they play the role of external sources for the CFT matter. Indeed, the holographic theory will remain fully non-linear in gravitational fields, what will be explicitly shown in Section 4. On the other hand, these theories couple successfully to external sources \cite{16}, which are stable in the framework of LCS supergravities \cite{17}.

The paper is organized as follows. In section 2 we introduce the holographic ansatz for the fundamental dynamical variables
and we arrive to their radial expansion in the asymptotic sector. Expressed in terms of the metric, it reduces to Fefferman-Graham expansion  \cite{18}. We also analyse corresponding residual gauge symmetries which leave this ansatz invariant.  In Section 3 we focus to the holographic quantum theory and derive the Noether-Ward identities. In section 4 we focus on Chern-Simons-AdS gravity in arbitrary odd  dimensions and compute 1-point functions in the corresponding dual theory, which are energy-momentum and spin currents. We show that translational and Lorentz symmetries are present also at the quantum level, but the Weyl anomaly and non-Abelian anomaly arise, breaking the conformal and non-Abelian symmetries quantically, the former being proportional to the Euler density up to a divergence. Our results generalize the ones of \cite{9} from five to arbitrary dimensions. Our calculations
are simplified to great extent by using the results of \cite{19}. Section 5 contains concluding remarks,
while appendices deal with some technical details.

Our conventions are given by the following rules. On a $D=d+1$ dimensional spacetime manifold $M$,
the Latin indices $(i,j,k,\dots)$ refer to the local Lorentz frame, the
Greek indices $(\m,\n,\r,\dots)$ refer to the coordinate frame. The  symmetric and antisymmetric parts of
a tensor $X_{ij}$ are $X_{(ij)}=\frac{1}{2}(X_{ij}+X_{ji})$ and
$X_{[ij]}=\frac{1}{2}(X_{ij}-X_{ji})$, respectively. The $d+1$
decomposition of spacetime is described in terms of the suitable
coordinates $x^\m=(\r,x^\a)$, where $\r$ is a radial coordinate and
$x^\a$ are local coordinates on the boundary $\pd M$. In the local
Lorentz frame, this decomposition is expressed by $i=(1,a)$.
\bigskip


\section{Holographic anzatz}

We are interested in a gravitational theory which possesses a local AdS symmetry. The presence
of local spacetime translations and spacetime rotations introduces naturally the vielbein
and the spin-connection as the fundamental fields. Our goal is to gauge fix this symmetry
by imposing a set of conditions on the fundamental fields in a such a way that it singles out
a particular coordinate frame which is suitable for a description of a holographically dual
theory. This frame should be consistent with the known Fefferman-Graham coordinate choice used
on the Riemannian manifold. All the properties that follow from this gauge-fixing are purely
kinematical and can be applied to any gravity invariant under local AdS group.
To include the dynamics we focus, in particular, on Lovelock-Chern-Simons gravity.

\subsection{AdS gauge transformations}
\setcounter{equation}{0}

In a theory with local AdS symmetry, the fundamental fields are components of a gauge
field (1-form) for the AdS group $SO(D-1,2)$ (see appendix A) and is defined by
\bea
A=\frac{1}{\ell}\,\he^A P_A+\frac 12\,\hom^{AB}J_{AB}\,,
\eea
where $\ell$ is the AdS radius. For the sake of simplicity, we set $\ell=1$.
Gauge transformations, parameterized  by $\l:=\eta^AP_A+\frac12 \l^{AB}J_{AB}$, act
on the gauge field as
\bea
\d_0A=D\l=d\l+[A,\l]\,,
\eea
wherefrom we get the transformation law of the fundamental fields,
\bea
&&\d_0\he^A=\hnab\eta^A-\l^{AB}\he_B\,,\nn\\
&&\d_0\hom^{AB}=\hnab\l^{AB}+2e^{[A}\eta^{B]}\,.
\eea
Here, the $\hom$-covariant derivative is
$
\hnab\eta^A:=d\eta^A+\hom^{AB}\eta_B\,.
$
The AdS field strength $F=dA+A\wedge A$ has components
\be
F=\hat T^AP_A+\frac12\,F^{AB}J_{AB}\,,
\ee
which are the torsion 2-form $\hat T^A$ and AdS curvature $F^{AB}$,
\bea
&&\hat T^A=\frac12\, \hat T^A{}_{\m\n}dx^\m \wedge dx^\n=d\he^A+\hom^{AB}\wedge\he_B\,,\nn\\
&&F^{AB}=\frac12\, F^{AB}{}_{\m\n}dx^\m \wedge dx^\n=d\hom^{AB}+\hom^{AC}\wedge\hom_C{^B}+\he^A\wedge\he^B\,.
\eea
The wedge product sign is going to be omitted for simplicity from now on in the text.
The global AdS space is described by a Riemannian manifold ($\hat T^A=0$), whose AdS curvature vanishes ($F^{AB} =0$),
and where the Riemannian curvature $\hat R^{AB}= d\hom^{AB}+\hom^{AC}\wedge\hom_C{^B}$ becomes
explicitly constant, $\hat R^{AB}=-\he^A\wedge\he^B$.

\subsection{Radial expansion and residual gauge transformations}

We use the radial foliation with the local coordinates $x^\mu=(\r,x^\a)$ and the Lorentz indices decomposed correspondingly as $A=(1,a)$. The asymptotic boundary of the manifold is located at the constant radius $\rho=\rho_0$. For convenience we set $\r_0=0$.

\prg{Gauge fixing.} There are two types of local symmetries, small and large, depending on how they behave asymptotically.
Small local symmetries are characterized by the parameters which go to zero at infinity and all other local symmetries are large. Small gauge symmetries act trivially on boundary fields and must be considered as redundancies in the theory, i.e., they must be gauge-fixed. A goodx
gauge choice should fix all small gauge transformations and should lead to a well-posed boundary value problem,
meaning that the form of a residual symmetry in the bulk is completely determined by the boundary values of the symmetry parameters. Note that the large gauge transformations do not have to be fixed by a gauge choice. For more details, see Ref. \cite{20}.

Local transformations at our disposal are spacetime diffeomorphisms and local AdS transformations.
Let us first focus on local AdS symmetry. A good gauge fixing for our purposes is the one where the spacetime is AAdS and where
residual gauge transformations contain conformal transformations on the boundary.

The last condition is introduced because we want to have a CFT as a holographic theory. Too strong gauge-fixing can overkill all residual transformations and give rise to a trivial holographic theory. Since the bulk theory is gauge invariant only up to boundary terms, different gauge-fixings can lead to non equivalent boundary theories.

Another important observation is that, in the metric formulation of Riemann gravity, according to the theorem of Fefferman-Graham (FG) \cite{18},
in any AAdS space there is a coordinate choice so that the metric can be cast in the FG form, that is, with $\hat g_{\r\r}=1/(2\r)^2$, $\hat g_{\r\a}=0$ and
$\rho \hat g_{\a\b}(\r,x)$ regular on the boundary $\r=0$. Thus, a gauge-fixing choice of the vielbein and spin-connection must be such that the corresponding metric acquires the FG form.

The number of gauge parameters of AdS group is $\frac{D(D+1)}2$, implying that we need
the same number of gauge conditions. We impose the following $D$ conditions on the vielbeins $\he^A{_\r}$ and $\frac{D(D-1)}2$ conditions
on connection $\hom^{AB}{}_\r$
\be
\he^A{_\r}=-\frac 1{2\r}\,\d_1{^A}\,,\qquad \hom^{AB}{}_\r=0\,.\lab{2.5}
\ee
In the choice of the gauge fixed one has to keep in mind the invertibility of vielbein. Therefore,
all $\he^A{_\r}$ components cannot be set to zero. Furthermore, although in principle a choice of the radial coordinate
is arbitrary, we want to have the Fefferman-Graham coordinate frame, where the metric component $g_{\rho \rho}$ behaves as $1/4\rho^2$, generalized to first order formalism, which implies the above behavior of the radial component of the vielbein.

To find residual transformations, we look at the restrictions on gauge parameters imposed by the gauge conditions \eq{2.5} and we find that they have to satisfy
\be
\begin{array}[b]{ll}
\pd_\r\eta^1=0\,,\medskip \quad  & \pd_\r\eta^a-\frac 1{2\r}\,\l^{1a}=0\,, \\
\pd_\r\l^{ab}=0\,, & \pd_\r\l^{1a}-\frac1{2\r}\,\eta^a=0\,.
\end{array}
\ee
The equations in $\eta^1$ and $\l^{ab}$ are straightforward to solve. To find $\eta^a$ and $\l^{1a}$, we combine the corresponding differential equations and obtain for the parameter $\eta^a$
\be
\r^2\pd_\r^2\eta^a+\r\pd_\r\eta^a-\frac14\,\eta^a=0\,. \lab{2.7}
\ee
This is the Euler-Fuchs equation which solution takes the form $\eta^a(\rho)\sim \r^k$. Hence, from \eq{2.7} we get
$k^2=\dis\frac14$ and consequently the general solution is given by
\begin{equation}
\begin{array}[b]{ll}
\eta^1(\r,x)=u(x)\,,\medskip & \eta^a(\r,x)=\frac1{\sqrt \r}\,\a^a(x)+\sqrt\r\,\b^a(x)\,, \\
\l^{ab}(\r,x)=\l^{ab}(x)\,,\quad & \l^{1a}(\r,x)=-\frac1{\sqrt \r}\,\a^a(x)+\sqrt\r\,\b^a(x)\,.
\end{array} \lab{2.8}
\end{equation}
We see that our gauge choice is good, as desired, because symmetry parameters in the whole bulk are determined by a few arbitrary functions $u$, $\a^a$, $\b^a$ and $\l^{ab}$ defined on the boundary.
We still have to identify an asymptotic symmetry group defined by these parameters.

The residual gauge parameters which describe asymptotic symmetry group naturally induce a change of the basis in the Lie
algebra $J_a^\pm=P_a\pm J_{1a}$, so that the Lie-algebra valued gauge parameter has
the form
\begin{equation}
\lambda =u(x)\,P_{1}+\frac{1}{\sqrt{\rho }}\,\alpha ^{a}(x)\,J_{a}^{-}
+\sqrt{\rho }\,\beta ^{a}(x)\,J_{a}^{+}+\frac{1}{2}\,\lambda ^{ab}(x)\,J_{ab}\,.
\end{equation}
The AdS algebra in terms of the new generators read
\begin{equation}
\begin{array}[b]{ll}
\left[ J_{a}^{+},J_{b}^{-}\right] =2J_{ab}+2\eta_{ab}P_1\,,\medskip   & \left[ J_{a}^{\pm },J_{b}^{\pm }\right]=0\,, \\
\left[ J_{ab},J_{c}^{\pm }\right] =-\eta_{ac}J_b^\pm+\eta_{bc}J_a^\pm\,,\medskip \quad & \left[ P_{1},J_{ab}\right]  =0\,, \\
\left[ P_{1},J_{a}^{\pm }\right] =\pm J_a^\pm\,. &
\end{array}
\end{equation}

\prg{Radial decomposition of gauge field and field strength.}
Up to now the  results are valid for any theory possessing AdS gauge symmetry. From now on we concentrate on Chern-Simons AdS gravity. For holography, one needs to know how the fields evolve along the radial direction and to study their near-boundary behavior. Since the radial
components are already fixed by the gauge condition \eq{2.5}, now we have to determine the behavior of the spatial components.

To this end, we can use invariance of gravity under general coordinate transformations. In Ref. \cite{21}, it was shown that only $D-1$ spatial diffeomorphisms are linearly independent on gauge generators, in a physical system where time-evolution was analyzed. In our case, we look at the radial quantization of a Hamiltonian, because we are interested in radial evolution of the fields from the bulk to the boundary. Thus, our independent diffeomorpisms act only in the transversal section of spacetime, that is, as $x^\a\rightarrow x^\a+\xi^\a(\r,x)$. Furthermore, we know that the radial diffeomorphisms are broken by the boundary set at constant radii, so this choice of quantization is natural in our case.

Thus, we have $D-1$ transversal diffeomorphisms to gauge-fix. In Ref. \cite{21} it was shown that, in any generic Chern-Simons gauge theory (AdS in our case), there is an on-shell identity $F_{\r\a}=F_{\a\b}\,N^\b$, with $D-1$ arbitrary functions $N^\b$ related to the transversal diffeomorphisms $\xi^\a(\r,x)$. Therefore, to gauge fix them, we can just set the $D-1$ functions to zero, $N^\b=0$. As a consequence, we also get $F_{\r\a}=0$  or, equivalently, $\hT^A{}_{\r\a}=F^{AB}{}_{\r\alpha}=0$. These conditions are particular for Chern-Simons theory and they arise from its dynamics. Interestingly, they can be exactly solved using the gauge fixing \eq{2.5}, also written as $A_\r =-\frac1{2\r}\, P_1$. Rewriting the AdS Lie-algebra valued condition $F_{\r\a}=0$ as $(dA+A^2)_{\r\a}=0$, we get
$$
\pd_\r A_\a-\frac1{2\r}\,\he^a{_\a}J_{a1}+\frac1{2\r}\,\hom^{1a}{}_\a P_A=0\,.
$$
This first order differential equation in $A_\a(\r,x)$ can be exactly solved, given the initial condition
\be
A_\a(0,x)\equiv e^a{_\a}(x)\,J_a^+ +k^a{_\a}(x)\,J_a^-+\frac{1}{2}\,\om^{ab}{_\a}(x)\,J_{ab}\,.
\ee
The solution is
\be
A_\a(\r,x)=\frac{1}{\sqrt{\r}}\,e^a{_\a}(x)\,J_{a}^{+}+\sqrt{\r}\,k^a{_\a}(x)\,J_{a}^{-}+\frac{1}{2}\,\om^{ab}{}_\a(x)\,J_{ab}\,.
\ee
In components, this solution leads to the radial expansion of the gravitational
fields expressed in terms of the boundary fields $e^a{_\a}$, $k^a{_\a}$ and $\omega^{ab}{_\a}$,
\bea
&&\he^{a}{_\a}=\frac1{\sqrt \r}(e^a{_\a}+\r k^a{_\a})\,,\nn\\
&&\hom^{1a}{_\a}=-\frac1{\sqrt \r}(e^a{_\a}-\r k^a{_\a})\,,\nn\\
&&\hom^{ab}{_\a}=\om^{ab}{_\a}\,.
\eea
Thus, this is a generalization of the FG expansion of the bulk metric. Indeed, the metric $\hat g_{\m\n}=\he^A{_\mu}\he^B{_\nu}\eta_{AB}$
takes the FG form since the line element can be written as
\be
ds^2=\frac1{4\r^2}\, d\r^2+\frac 1\r\left(g_{\a\b}+2\r k_{(\a\b)}+\r^2k^a{_\a}k_{a\b}\right)dx^\a dx^\b\,,
\ee
where $g_{\a\b}:=\eta^{ab}e^a{_\a}e^b{_\b}$ and $k_{\a\b}:=e_{a\a}k^a{_\b}$.
We conclude that the FG expansion is {\it finite}. Finite FG expansion is typical for Chern-Simons gravity \cite{9} and also for General Relativity when the Weyl tensor vanishes \cite{8}.

The induced metric $\g_{\a\b}$ is defined by $\g_{\a\b}=\r \hat g_{\a\b}$.
The coefficients in the radial expansion of $\g_{\a\b}$ are
\be
\begin{array}[b]{ll}
\g_{\a\b}^{(0)}=g_{\a\b}\,,\medskip & \g_{\a\b}^{(1)}=2k_{(\a\b)}\,, \\
\g_{\a\b}^{(2)}=k^a{_\a}k_{a\b}\,, \quad & \g_{\a\b}^{(n)}=0\,,\;n\geq 3\,.
\end{array}
\ee

From the radial expansion of the field-strength we get on the boundary
\be
\begin{array}[b]{ll}
F^{a1}=\frac 1{\sqrt \r}\left(T^a-\r\nab k^a\right)\,,\medskip \quad  & \hat T^1=-2e^ak_a\,, \\
F^{ab}=R^{ab}+4e^{[a}k^{b]}\,,& \hat T^a=\frac1{\sqrt\r}(T^a+\r\nab k^a)\,,
\end{array}
\ee
where $T^a=\nab e^a$ and $R^{ab}=d\om^{ab}+\om^a{_c}\om^{cb}$.

Physical interpretation of the boundary fields can be found from their transformation law under the residual (boundary) gauge transformations.

\prg{Residual gauge transformations.} The complete transformation law of the basic dynamical variables in the bulk that include the spacetime diffeomorphisms is given by
\bea
&&\d_0\he^A{_\mu}=\hat\nab_\mu\eta^A-\l^{AB}\he_{B\mu}-\pd_\mu\xi^\nu\he^A{_\nu}-\xi^\nu\pd_\nu\he^A{_\m}\nn\,,\\
&&\d_0\hom^{AB}{_\mu}=\hat\nab_\mu\l^{AB}+2\he^{[A}{_\mu}\eta^{B]}-\pd_\mu\xi^\nu\hom^{AB}{_\nu}-\xi^\nu\pd_\n\hom^{AB}{_\m}\,,\lab{2.15}
\eea
where the last two terms of each line are the Lie derivatives with respect to $\xi^\mu$. If we make the following redefinition of parameters,
\bea
&&\eta^A\rightarrow\eta^A+\xi^\mu\he^A{_\mu}\,,\nn\\
&&\l^{AB}\ra\l^{AB}+\xi^\mu\hom^{AB}{_\mu}\,,\lab{2.16}
\eea
transformations \eq{2.15} take the following form,
\bea
&&\d_0\he^A{_\mu}=\hat\nab_\mu\eta^A-\l^{AB}\he_{B\mu}+\xi^\nu \hat T^A{}_{\m\n}\nn\,,\\
&&\d_0\hom^{AB}{_\mu}=\hat\nab_\mu\l^{AB}+2\he^{[A}{_\mu}\eta^{B]}+\xi^\nu F^{AB}{}_{\m\n}\,.\lab{2.17}
\eea
Due to the condition $F_{\r\a}=0$, the transformation laws \eq{2.17} of  $\he^A{_\r}$ and $\hom^{AB}{}_\r$ with redefined parameters \eq{2.16} take the {\it same form}
as in the case when diffeomorphisms are absent in the transformation law \eq{2.15}. Therefore, introduction of diffeomorphisms
\emph{does not effectively change} the result \eq{2.8}.

From the transformation law for $\om^{ab}{_\a}$, it follows that $\xi^\a$ \emph{\emph{does not depend on }}$\r$.
The complete transformation law of the gauge fields under residual transformations read
\bea\lab{2.19}
&&\d_0e^a{_\a}=\nab_\a\a^a-\l^{ab}e_{b\a}+ue^a{_\a}-\xi^\b{}_{,\a}e^a{_\b}-\xi^\b\pd_\b e^a{_\a}\,,\nn\\
&&\d_0k^a{_\a}=\nab_\a\b^a-\l^{ab}k_{b\a}-uk^a{_\a}-\xi^\b{}_{,\a}k^a{_\b}-\xi^\b\pd_\b k^a{_\a}\,,\nn\\
&&\d_0\om^{ab}{_\a}=\nab_\a\l^{ab}+4e^{[a}{_\a}\b^{b]}+4k^{[a}{_\a}\a^{b]}-\xi^\b{}_{,\a}\om^{ab}{_\b}-\xi^\b\pd_\b \om^{ab}{_\a}\,,
\eea
with
\be
\eta^1+\frac{\xi^\r}{2\r}=u(x)\,,\qquad
\xi^\a=\xi^\a(x)\,.
\ee
Let us note that the residual diffeomorphisms do not change the condition $F_{\r\a}=0$, as expected. Their form
shows that our gauge choice is good.

In holography it is important the boundary to be orthogonal to the radial direction. That is why
we shall impose an additional condition $\he^1{_\a}=0$, which puts the bulk vielbein in the block-diagonal form with the only one boundary component $e^a{_\a}(x)$. The extra condition reduces the asymptotic symmetries because the parameter $\b^a$ is not independent any longer,
\bea
\b^a=e^{a\a}\left(\frac12\,\pd_\a u+k^b{_\a}\a_b\right)\,.
\eea
The generators of the asymptotic group cannot be determined
straightforwardly because a change of the basis of the Lie algebra necessary to identify this subgroup is non-linear, that is, it depends on the point of spacetime. We shall deduce the algebra directly from the action on the fields.

Independent transformations acting on the fields are transversal diffeomorphisms or local translations $\d_T(\xi)$, local Lorentz rotations $\d_L(\l)$, local Weyl or conformal transformations $\d_C(u)$ and non-Abelian gauge transformations $\d_G(\a)$. Each transformation can be seen as generated by some generator $T_a$ through the commutator, for example $[\d_G(\a'),\d_G(\a'')]=\a'^a\a''^b[T_a,T_b]$, and similarly for all other transformations. In that way, the asymptotic algebra closes as
\be
\begin{array}[b]{ll}
\left[\d_T(\xi'),\d_T(\xi'')\right]=\d_T([\xi',\xi''])\,,\medskip \qquad  & \left[\d_C(u),\d_G(\a)\right]=\d_C(\a\cdot\pd u)-\d_L(\tilde{\l})-\d_G(u\a)\,, \\
\left[\d_T(\xi),\d_L(\l)\right]=\d_L(\xi\cdot\pd\l)\,,\medskip  & \left[\d_G(\a'),\d_G(\a'')\right]=-\d_C(\tilde{u})-\d_L(\Lambda)\,, \\
\left[\d_T(\xi),\d_C(u)\right]=\d_C(\xi\cdot\pd u)\,,\medskip  & \left[\d_L(\l,\d_G(\a)\right]=\d_G(\l\cdot\a)\,, \\
\left[\d_T(\xi),\d_G(\a)\right]=\d_G(\xi\cdot\pd\a)\,,\medskip  & \left[\d_L(\l),\d_C(u)\right]=0\,, \\
\left[\d_L(\l'),\d_L(\l'')\right]=\d_L([\lambda',\lambda''])\,, & \left[\d_C(u'),\d_C(u'')\right]=0\,,
\end{array}\lab{2.21}
\ee
where $[\xi',\xi'']^\a=\xi'\cdot\pd\xi''^\a-\xi'\cdot\pd\xi''^\a$ is the Lie bracket and
$[\lambda',\lambda'']^{ab}=\l'^{ac}\l''_c{}^b-\l''^{ac}\l'_c{}^b$ is the group commutator. We also introduced the
contraction $\xi \cdot \partial =\xi ^{\beta }\partial _{\beta }$ and the matrix
multiplication $(\l\cdot\a)^a=\l^{ab}\a_b$, and defined the auxiliary Lorentz parameters $\tilde{\l}^{ab}=2\a^{[a}\pd^{b]}u$
and $\Lambda ^{ab}=4k^{c[a}(\a'_c\a''^{b]}-\a''_c\a'^{b]})$, as well as the Weyl parameter $\tilde{u}=4k^{[ab]}\a'_a\a''_b$.

The above brackets are computed by acting on $e^a{_\a}$, but their form is field-independent. The boundary diffeomorphisms, Lorentz rotations and Weyl dilatations close in the standard way and they form the Weyl subgroup. Furthermore, the non-Abelian extension is realized non linearly, because the parameters $\Lambda$ and $\tilde{u}$ explicitly depend on the field $k^{ab}$.  To understand better the origin of such non-Abelian transformations, let us note that
\be
\d_G(\a)e^a{_\a}=\left(\pd_\a\a^\b\right)e^a{_\b}+\a^\b\pd_\b e^a{_\a}+\a^\b\om^{ab}e_{b\a}+\a^\b T^a{}_{\a\b}\,,\lab{2.18}
\ee
where $\a^\b=\a^a e_a{^\b}$. Therefore, the gauge transformations can be cast in the form
\be
\d_G(\a)e^a{_\a}=-\d_T(\a^\b)-\d_L(\om^{ab}{}_\b\a^\b)+\a^\b T^a{}_{\a\b}\,.
\ee
Shifting the parameters as $\xi^\b \ra \xi^\b+\a^\b $ and
$\lambda^{ab}\rightarrow \lambda ^{ab}+\omega^{ab}{}_\beta \alpha ^\beta $
helps us identify the independent non-Abelian gauge transformations
$\delta_G(\alpha)e^a{_{\alpha }}=\alpha ^\beta T^{a}_{\alpha \beta }$.
From \eq{2.18} and the above relation we easily conclude that  non-Abelian gauge transformations
act on the boundary vielbein independently if and only if torsion is non-vanishing.
In the case of vanishing torsion non-Abelian gauge transformations stop to be independent and they can be represented as composition of local translations and local  Lorentz rotations with the suitable redefinition of parameters. Similar conclusion holds when one acts on the boundary spin connection
because it is an independent field only if the torsion is non-vanishing.

Let us now, for completeness, inspect the action of  the transformations \eq{2.19} on the metric $g_{\a\b}=e^a{_\a}e_{a\b}$. We obtain
$$
\d_0 g_{\a\b}=-\xi^\g{}_{,\a}g_{\g\b}-\xi^\g{}_{,\b}g_{\a\g}-\xi^\g\pd_\g g_{\a\b}+2ug_{\a\b}+e_{a\b}\nab_{\a}\a^a+e_{\a\a}\nab_{\b}\a^\a\,.
$$
Similarly, as in the  the case of vielbein, the action of the non-Abelian gauge transformations on the metric
reads
\be
\d_G(\a) g_{\a\b}=-\d_T(\a)g_{\a\b}+2\a^\g T_{(\a\b)\g}\,,
\ee
Again, we conclude that in the case when torsion vanishes the action of non-Abelian gauge transformations on the metric reduces to local translations with the already mentioned redefinition of parameters \cite{4}. The above transformation law of the metric is not usual in field theories, but is not surprising because we started with local AdS symmetry which mixes vielbein and spin connection.

\section{Noether-Ward identities}
\setcounter{equation}{0}

The AdS/CFT correspondence between the $D$-dimensional AdS space and $d$-dimensional CFT identifies the quantum effective action in CFT with the classical gravitational action in AdS space for given boundary conditions. Thus, let us assume that the {\it renormalised} effective action in a holografic theory, $I_{\rm ren}[e,\om]$, has an extremum for Dirichlet boundary conditions on the independent fields, which are the vielbein, $e^a{_\a}$, and the spin-connection, $\om^{ab}{_\a}$, so that its variation takes the form
\bsubeq\lab{3.1}
\be
\d I_{\rm ren}[e,\om]=-\int d^dx\left(\t^\a{_a}\d_0e^a{_\a}+\frac12\s^\a{}_{ab}\d_0\om^{ab}{_\a}\right)\,.
\ee
The tensor densities,
\be
\t^\a{_a}=-\frac{\d I_{\rm ren}}{\d e^a{_\a}}\,,\quad \s^\a{}_{ab}=-\frac{\d I_{\rm ren}}{\d \om^{ab}{_\a}},
\ee
\esubeq
are the energy-momentum and spin currents of our dynamical system.

The holographic theory is invariant under $d$-dimensional diffeomorphisms with the parameter $\xi^\a$ and  the local Lorentz transformations with the parameter $\l^{ab}$. The conservation law of the corresponding Noether current reads
\bsubeq\lab{3.2}
\bea
&&e^a{_\b}\nab_\a\t^\a{_a}+\t^\a{_a}T^a{}_{\a\b}+\frac12\s^\a{}_{ab}R^{ab}{}_{\a\b}+\frac 12\om^{ab}{}_\b(\nab_\a\s^\a{}_{ab}-2\t_{[ab]})=0\,,\lab{3.2a}\\
&&\nab_\a\s^\a{}_{ab}-2\t_{[ab]}=0\,,\lab{3.2b}
\eea
which is also known as the generalized conservation laws of $\t_\a{^a}$ and $\s^\a{}_{ab}$. Note that if the
second Noether identity \eq{3.2b} is fulfilled, the last term in \eq{3.2a} can be omitted.
We shall keep this term, however, because it modifies the conservation law in case when there are quantum anomalies.

The invariance of $I_{\rm ren}$ under Weyl transformations leads to the additional conservation law,
\be
\t-\nab_\b\s^a{_a}{}^\b=0\,,\lab{3.2c}
\ee
where $\tau:= \tau^a{_a}$  is the trace of the energy-momentum tensor.

Finally, invariance under the non-Abelian gauge transformations leads to
\be
\nab_\a\t^\a{_a}-2\s^b{}_{bc}k_a{^c}-2\s_{bca}k^{cb}=0\,.\lab{3.2d}
\ee
\esubeq
In Ref. \cite{9}, it was proposed that these residual gauge transformations contain the information about the chiral anomaly of the fermions in holographic CFT, encoded in the completely antisymmetric part of the spin current.

Gravitational dynamics in the bulk is described by non-vanishing torsion, but
it may happen that some solutions on the boundary are Riemannian.  For such
solutions, the boundary connection $\om^{ab}{_\a}$  takes its Riemannian value $\tom^{ab}{_\a}=\tom^{ab}{_\a}(e)$ and can be expressed in terms of the
the vielbein $e^a{_\a}$ in the following way,
\be
\tom_{ab\a}=\frac12\left(c_{abc}-c_{cab}+c_{bca}\right)e^c{_\a}\,,\quad c_{a\a\b}:=\pd_\a e_{a\b}-\pd_\b e_{a\a}\,.
\ee
Although boundary connection is no more independent dynamical variable,  the Noether--Ward identities keep the form \eq{3.2}, but now
$\om_{ab\a}$ takes on the Riemannian value $\tom_{ab\a}$.

From the Riemannian renormalised action $\tilde I_{\rm ren}=I_{\rm ren}[e^a{_\a},\tom_\a]$, we get that the related spin
current $\S^\a:=-\d\tilde I_{\rm ren}/\d\om_\a$ vanishes, while the energy-momentum current $\Th^\a{_a}:=-{\d\tilde I_{\rm ren}}/{\d e^a{_\a}}$
acquires an additional contribution
\be
\Th^\a{_a}=\tilde\t^\a{_a}
   -\frac 12\tilde\nab_\b\left(\tilde\s^{\b\a}{}_{a}-\tilde\s_a{}^{\b\a}+\tilde\s^\a{}_a{}^\b\right)\,,
\ee
where $\tilde X$ denotes the Riemannian limit of a tensor $X$.
The Noether identities for the action $\tilde I_{\rm ren}$ are found to be
\bsubeq\lab{3.5}
\bea
&&e^a{_\b}\tilde\nab_\a\Th^\a{_a}-\tom^{ab}{}_\b\Th_{[ab]}=0\,,\lab{3.5a}\\
&&\Th_{ab}=\Th_{ba}\, , \lab{3.5b}                                    \\
&&\Th=0\, . \lab{3.5c}
\eea
\esubeq
Let us remind that, as we concluded at the end of the previous section, the
non-Abelian gauge transformations are not independent for Riemannian  solutions, thus
in this case there are only three  independent Noether identities \eq{3.5}.

When the Lorentz invariance is fulfilled, \eq{3.5a} reduces to the usual form
$D_\a(e^{-1}\Th^\a{_\b})=0$, where $D_\a$ is the Riemannian covariant
derivative. The relations \eq{3.5b} and \eq{3.5c} are the standard Riemannian
conditions for the Lorentz and Weyl invariance, respectively.

After using the condition of vanishing torsion, $T_{abc}=0$, the identity
$[\tilde\nab_\a,\tilde\nabla_\b] f_a
  =\tilde R_{ab\a\b}f^b$ and the Bianchi identity, $\tilde R_{abcd}+\tilde R_{acdb}+\tilde R_{adbc}=0$, enable to write the expressions \eq{3.5} as
\bsubeq
\bea
&&e^a{_\b}\tilde\nab_\a\tilde\t^\a{_a}+\frac12\tilde\s^\a{}_{ab} \tilde R^{ab}{}_{\a\b}
      +\frac 12\tom^{ab}{}_\b(\nab_\a\tilde\s^\a{}_{ab}-2\tilde\t_{[ab]})=0\,,\\
&&\tilde\nab_\a\tilde \s^\a{}_{ab}-2\tilde\t_{[ab]}=0\,,   \\
&&\tilde\t-\tilde\nab_\b\tilde\s^a{_a}{}^\b=0\,.
\eea
\esubeq
Hence, the Riemannian identities \eq{3.5a}, \eq{3.5b} and \eq{3.5c} coincide with those obtained
from \eq{3.2a}, \eq{3.2b} and \eq{3.2c} in the limit $T_{abc}\to 0$, as expected. Therefore, taking
torsionless limit and calculating Noether-Ward identities gives an equivalent result
as first calculating the Ward identities and taking torsion zero \cite{22}. This is important when we do not
know whether the torsion vanishes. Therefore, one may safely work in first order formalism assuming the boundary conditions and
gauge fixing presented previously.

\section{Lovelock-Chern-Simons gravity}
\setcounter{equation}{0}
\subsection{Action and equations of motion}

The Lovelock-Lanczos gravity \cite{15} in  first order formulation  is described by the action
\bsubeq
\be
I_{\rm L}=\sum_{p=0}^{[D/2]}\a_p L_p\,,
\ee
where $\a_p$ are arbitrary coupling constants and  $L_p$ is dimensionally continued Euler density in $D$ dimensions,
\be
L_p=\ve_{i_1 i_2 \dots i_D} R^{i_1i_2} \dots R^{i_{2p-1} i_{2p}}e^{i_{2p+1}} \dots e^{i_D}.
\ee
\esubeq
Here  $p$ is the power of the curvature tensor in the polynomial $L_p$. We omit writing the wedge product for the sake of simplicity.

Lovelock-Lanczos gravity possesses numerous black hole solutions with Riemannean geometry \cite{23,24,25}, although some choices of the coupling constants $\{\a_p\}$ exhibit a causality problem in the dual CFT \cite{26}, or have instable geometries \cite{27,28}. Generic Lovelock gravity without torsion possesses the same number of degrees of freedom as general relativity \cite{{29}}. With torsion, or when the parameters take the critical values, the dynamical content of Lovelock-Lanczos gravity might change. Solutions in these cases are known as well, for example the ones with Riemann-Cartan geometry in five dimensional gravity \cite{30,31} and supergravity \cite{32}.

In odd-dimensional case $D=2n+1$, the  special choice of coefficients  $\a_p=\frac{\k}{2n+1-2p}\binom{n}{p}$ defines theory with the {\it unique} (degenerate) AdS vacuum, known as Lovelock Chern-Simons (LCS) AdS gravity. Alternatively, LCS action can be constructed as a Chern density by taking the topological invariant, Chern form
 $dL_{\rm CS}=\ve_{i_1j_1\dots i_nj_n} F^{i_1j_1}\dots F^{i_nj_n}$, and writing $L_{\rm CS}$ by using holonomy
operator \cite{14,33}. Then, an equivalent form of LCS action is given by
\be
I_{\rm LCS}=\k\int\limits_\cM\int\limits_0^1dt\,\ve_{A_1B_1A_2B_2\dots A_nB_nC}\prod_{k=1}^{n}\left(\hR^{A_kB_k}+t^2\he^{A_k}\he^{B_k}\right)\he^C\,. \lab{4.1}
\ee
Dropping the indices for simplicity, the above expression reads
\bea
I_{\rm LCS}=\k\int\limits_\cM\int\limits_0^1dt\,\ve\left(\hR+t^2\he^2\right)^n\he=\k\int\limits_{\cM}\sum_{k=0}^n\binom{n}{k}\frac1{2k+1}\,\ve\hR^{n-k}\he^{2k+1}\,,\lab{4.2}
\eea
where we used the binomial expansion to perform an integration over $t$.

Equations of motion are obtained from the variation of the action \eq{4.2} with respect to fundamental variables $\he^A$ and $\hom^{AB}$.
Variation with respect to $\he$ yields
\bea
C_A:=\ve_{AA_1B_1\dots A_nB_n}\prod_{k=1}^{n}F^{A_kB_k}=0\,,\lab{4.3}
\eea
which can be split into $1$ and $a$ components,
\bsubeq
\bea
&&C:=\ve_{1a_1b_1\dots a_nb_n}\prod_{k=1}^{n}F^{a_kb_k}=0\,,\lab{4.4a}\\
&&C_a:=\ve_{a1ba_2b_2\dots a_nb_n}F^{1b}\prod_{k=2}^{n}F^{a_kb_k}=0\,.\lab{4.4b}
\eea
\esubeq
Variation with respect to $\om$ yields
\bea
C_{AB}:=\ve_{ABA_1B_1\dots A_{n-1}B_{n-1}C}\prod_{k=1}^{n-1}F^{A_kB_k}\hT^C=0\,,\lab{4.5}
\eea
and can be split into $[1a]$ and $[ab]$ components,
\bsubeq\lab{4.6}
\bea
&&{\bar C}_a:=\ve_{1aa_1b_1\dots a_{n-1}b_{n-1}c}\prod_{k=1}^{n-1}F^{a_kb_k}\hT^c=0\,,\lab{4.5a}\\
&&C_{ab}:=\ve_{1aba_1b_1\dots a_{n-1}b_{n-1}}\prod_{k=1}^{n-2}F^{a_kb_k}\left(F^{a_{n-1}b_{n-1}}\hT^1+(n-1)F^{1a_{n-1}}\hT^{b_{n-1}}\right)\,.
\eea
\esubeq

Let us note that $\hat T^a=0$ is a particular solution of the equations \eq{4.6} belonging to the  Riemannian sub-class of all solutions of the theory.
Also, the global AdS space ($F^{ab}=0$) is a particular solution of all equations of motion.

\subsection{1-point functions}

In this section we calculate the renormalized gravitational LCS action in the classical approximation. Then we use the AdS/CFT correspondence to promote it to the quantum effective action in a holographic CFT, and compute the holographic 1-point functions.

The variation of the LCS action reads
\bea
\d I_{\rm LCS}=n\k\int\limits_{\pd\cM}\int\limits_0^1dt \ve_{ABCA_1B_1\dots A_{n-1}B_{n-1}}\d\hom^{AB}\he^C\prod_{k=1}^{n-1}\left(\hR^{A_kB_k}+t^2\he^{A_k}\he^{B_k}\right)\,.
\eea
To perform a near-boundary expansion of the fields, let us first rewrite the following quantity in terms of the AdS tensor,
$$
\hR^{A_kB_K}+t^2\he^{A_k}\he^{B_k}=F^{A_kB_k}+(t^2-1)\he^{A_k}\he^{B_k}\,.
$$
the first term in the above expression is {\it independent} of $\r$ since on the boundary $\he^1=0$, and therefore
the particular components expand as
\bea
\hR^{a_kb_k}+\he^{a_k}\he^{a_k}&=&F^{a_kb_k}\,,\nn\\
\hR^{a_k1}+\he^{a_k}\he^1&=&\frac1{\sqrt\r}\left(T^{a_k}-\r\nab k^{a_k}\right)\,.
\eea
Plugging these expansions in the variation of the action, we find
\bea
\d I_{\rm LCS}=n\k\int\limits_{\pd\cM}\ve\,\d\hom\sum_{k=0}^{n-1}\binom{n-1}{k}\frac{(-1)^k(2k)!!} {(2k+1)!!}\,F^{n-k-1}\he^{2k+1}\,,\lab{4.7}
\eea
where we used the beta function to solve the integral $\int_0{^1}dt(t^2-1)^k=\dis\frac{(-1)^k(2k)!!} {(2k+1)!!}$.

Variation \eq{4.7} is divergent on the boundary, that is, in the limit $\r\ra 0$ and extraction of physical quantities requires its renormalisation, or removal of divergences. For related work on Riemannian Lovelock gravity, see Ref. \cite{5}.

The procedure for obtaining finite results consists in introducing a regulating surface at $\r=\epsilon$ and adding the counterterms which cancel all divergent contributions as $\epsilon$ tends to zero \cite{8,34}. Equivalently said, the divergent terms in a variation of an action have to be represented as total variations of local terms integrated over boundary.
In general, the computation of the total variation can be substantially simplified after noting that the conditions for the application of the theorem \cite{19}  are fulfilled in our case.
For an alternative proof of the theorem \cite{19}, see appendix C. The theorem \cite{19} states  that the terms which are asymptotically divergent or zero (when $\r\ra 0$) can always be represented as total variations of local boundary functionals. Therefore, we can discard all $\rho^{\a}$ ($\a\neq 0$) terms in the expression \eq{4.7} and keep only the $\r^0$-terms. For
the form of the $\r^\a$-terms ($\a\neq 0$), see appendix B. Note that the counterterms can contain arbitrary local finite part which is non-physical and depends on a renormalisation scheme. The divergent counterterms are local and there is finite number of them. They also depend on only one coupling constant $\k$.
Counterterms in Riemannian gemetry were calculated in Ref. \cite{35}.

Keeping only the finite terms, we obtain the variation of the regularized action $I_{\rm ren}=I_{\rm LCS}+I_{\rm ct}$ in the form
\bea
\d I_{\rm ren}&=& -2n\k\ve\left[\d \om T\sum\limits^{n-2}_{l=0}\binom{n-2}{l}\frac{(-1)^l 2^{2l+1}(n-1)}{l+1}\,(R+4ek)^{n-2-l}e^lk^{l+1}\right.\nn\\
&&-\left.\delta e \sum\limits^{n-1}_{l=0}\binom{n-1}{l}\frac{(-1)^l 2^{2l+1}}{l+1}\,(R+4ek)^{n-1-l}e^lk^{l+1}\right]\,,
\eea
where $T=\nabla e$ is the boundary torsion tensor. Comparing to \eq{3.1}, the spin and energy-momentum  currents are given by, respectively,
\bea
&&\s_{ab}=-n\k\,\ve_{1ab}\,T\sum\limits^{n-1}_{l=1}\binom{n-1}{l}4^l R^{n-1-l}e^{l-1}k^l\,,\label{sigma}\\
&&\t_a=\k\,\ve_{1a}\sum\limits^{n}_{l=1}\binom{n}{l}4^lR^{n-l}e^{l-1}k^l\,,\label{tau}
\eea
and they correspond to the vevs of the quantum CFT operators, the spin-current ${\cal S}_{ab}$ and the energy-momentum of the conformal matter
${\cal T}_{a}$,
\be
\s_{ab}=\left<{\cal S}_{ab}\right>_{\rm CFT}\,,\qquad \t_a=\left<{\cal T}_{a}\right>_{\rm CFT}\,.
\ee
Using these representations of the 1-point functions of the CFT operators, we can study their quantum conservation laws, that is, the Noether-Ward identities.

\subsection{Anomalies}

The equations \eq{3.2} describe classical conservation laws in a holographic theory invariant under diffeomorphisms, conformal transformations and non-Abelian gauge transformations. Since now we know the form of the corresponding quantum currents, we can also check the quantum conservation laws. If the law is not satisfied, then the quantum theory possesses a quantum anomaly.

In this section we explore the Ward identities and check for the existence of quantum anomalies: Lorentz anomaly $A_{ab}$, diffeomorphism anomaly $\bar A_a$, conformal anomaly $A$ and
gauge anomaly $A_a$. It is well-known that there are two types of non-Abelian anomalies, covariant and consistent. All the anomalies we compute here are covariant, i.e., they transform covariantly under gauge symmetries.

\prg{Lorentz Ward identity.} The conservation law for Lorentz symmetry is given by Eq.\eq{3.2b}, so we have to calculate the quantity
\bea
A_{ab}=\nab\s_{ab}-2e_{[a}\t_{b]}\,.
\eea
Using the expressions (\ref{sigma}) and (\ref{tau}) for the quantum currents, we find
\bea
A_{ab}&=&-4n\k \ve_{ab}\left[2(n-1)T\nabla k\sum\limits^{n-2}_{l=0}\sum\limits^{n-2-l}_{m=0}\binom{n-2}{l}\binom{n-2-l}{m}4^m(l+m+1)R^{n-2-l-m}e^{l+m}k^{l+m}\right.\nn\\
&&+\left.e_c k^c\sum\limits^{n-1}_{l=0}\sum\limits^{n-1-l}_{m=0}\binom{n-1}{l}\binom{n-1-l}{m}\frac{(-1)^l2^{2l+2m+1}(l+m+1)}{l+1}\,R^{n-1-l-m}e^{l+m}k^{l+m}]\right].\nn
\eea
It turns out that $A_{ab}$ can be completely expressed in terms of the field equations, that means that it vanishes,
\bea
A_{ab}=-4n\k \,C_{ab}= 0\,.
\eea
Therefore, there is no Lorentz anomaly in the holographic theory because the Lorentz symmetry is conserved also quantically. This is an expected result, since the Lorentz symmetry is usually broken in the actions that are not parity invariant.

\prg{Ward identity for diffeomorphisms.} The conservation law for local translations has the form \eq{3.2a},
\bea
\bar A_a=\nab \t_a - (I_aT^b\t_b+\frac{1}{2}I_aR^{bc}\s_{bc})\,,
\eea
where $I_a$ is the contraction operator with the spacetime index projected to the tangent manifold using the inverse vielbein ${e_a}^\a$. Plugging in the quantum currents (\ref{sigma}) and (\ref{tau}), one can show that the conservation law is satisfied,
\bea
\bar A_a=4n\kappa\,(k^{b}_{\ a}C_b-\bar{C}_a) = 0\,.
\eea
Therefore, there is no gravitational anomaly, as expected.

\prg{Conformal Ward identity.} The conservation law for local Weyl transformations can be read off from Eq. \eq{3.2c}  as
\bea
A=e^a\tau_a+\nabla(e^aI^b\sigma_{ab})\,,
\eea
where $e^a\tau_a$ is the trace of energy-momentum tensor, so $A$ is also called the trace anomaly. Using the field equations and discarding the total divergence, one can show that the trace anomaly has the form
\bea
e^a\tau_a=\kappa\, \ve_{a_1b_1 a_2b_2\ ...\ a_n b_n}R^{a_1b_1}R^{a_2b_2}...R^{a_n b_n}=\kappa\, {\cal E}_n(R)\,.
\eea
Thus, the holographic anomaly is non-vanishing and, up to a divergence,  proportional to the  Euler density ${\cal E}_n(R)=\ve R^n$,
as expected in a CFT dual to a higher-dimensional AdS gravity \cite{36}.
Since the Weyl anomaly is topological invariant, it is of the type A, according to the general
classification of conformal anomalies given in Ref.  \cite{37}.

\prg{Ward identity for gauge symmetry.}   The conservation law for non-Abelian gauge transformations is given by Eq. \eq{3.2d} as
\be
A_a=\nab \t_a - 2(e^b \s_{bc}k^{\ c}_a+k^b\s_{ba})\,.
\ee
Using (\ref{sigma}) and (\ref{tau}), as well as the equations of motion,  we can express it as
\be
\begin{aligned}
A_a=
&-2n\k\,\ve_{1b}\,I_aT^b\sum\limits^{n-1}_{l=0}\binom{n-1}{l}\frac{(-1)^l2^{2l+1}}{l+1}\,(R+4ek)^{n-1-l}e^lk^{l+1}\\
&-4n\k\,\ve_{1bc}T\left(\frac{1}{2}\,I_a R^{bc}-2e^bk^{\ c}_a\right)\sum\limits^{n-2}_{l=0}\binom{n-2}{l}\frac{(-1)^l2^{2l+1}(n-1)}{l+1}\,(R+4ek)^{n-2-l}e^lk^{l+1}\\
 &+8n\k\ve_{1a}T\sum\limits^{n-2}_{l=0}\binom{n-2}{l}\frac{(-1)^l2^{2l+1}(n-1)}{l+1}\,(R+4ek)^{n-2-l}e^lk^{l+2}\neq 0\,.
\end{aligned}
\ee
The above holographic anomaly is in general non vanishing, but it cancels out when \emph{ the torsion is equal to zero,} as expected. Indeed, when $T^a=0$, the non-Abelian gauge symmetry
 is not independent, but it can be expressed in terms of the diffeomorphisms, which are conserved at the quantum level.
Another derivation of this result is possible by noting that in this particular case the spin tensor vanishes and both Eqs. \eq{3.2a} and \eq{3.2d} reduce to
\be
\tilde\nab_\a\tilde\t^\a{_\a}=0\,.
\ee
Again non-Abelian gauge anomaly vanishes since $A_a=0$.

\section{Concluding remarks}

We analyzed a holographic dual of Lovelock Chern-Simons AdS gravity in an arbitrary odd dimension and calculated corresponding holographic currents and anomalies in the quantum CFT. First part of the work is devoted to the kinematics of gravitational theory with AdS gauge symmetry. After motivating a gauge fixing suitable for a holographc analysis, we calculated residual (asymptotic) symmetries. Then we focused to Chern-Simons AdS gravity. We concluded  that
 the largest asymptotic symmetry consists of local translations and rotations (local Poincar\'e group),  local Weyl rescalings and, in the presence of torsion on the boundary, of non-Abelian gauge symmetry.
If the torsion on the boundary is zero, then a non-Abelian symmetry is not independent any longer and reduces to local Poincar\'e transformations.

We found holographic representations of the energy-momentum and spin tensors in a dual theory, which we identified with the corresponding
1-point functions in CFT, in the presence of sources. We also computed their conservation laws and obtained that some of quantum symmetries are broken, leading to quantum anomalies. Explicitly, we
obtained that local translations and rotations are symmetries of the quantum theory, while Weyl rescalings and non-Abelian gauge symmetry are anomalous.
Similarly as in five dimensions \cite{9}, the trace anomaly is proportional to the Euler density and is therefore of the type A.

Because of  non-linearity of the model and working in higher-dimensional Riemann-Cartan space, the regularization of the action was quite involved.  However, with the help of a general renormalization theorem shown in appendix C,  it was possible to circumvent an explicit construction of divergent counterterms and extract directly its finite part. An alternative proof of the theorem is given in Ref. \cite{19}.

One of the open questions left for  future work is an application on non-Abelian gauge transformations to the calculation of chiral anomaly. Namely, in  Ref. \cite{9} it was suggested that the chiral anomaly is related to the completely antisymmetric component of the torsion tensor. Another question would be to find a different gauge fixing of either transversal diffeomorphisms or local AdS symmetry, in order to obtain an {\it infinite} radial expansion of the fields, and possibly the type B anomaly. This would describe an inequivalent holographic theory. Finally, we are also interested in introducing a gauge-fixing which breaks relativistic covariance in an arbitrary Poincar\'e gauge theory, and is suitable for the formulation of Lifshitz holography. This last topics is the work in progress.

\section*{Acknowledgments}
The autors thank Milutin Blagojevi\'c for many useful discussions and comments.
This work was partially supported by the Serbian Science Foundation under Grant No. 171031, Chilean FONDECYT Project
No.1170765 and the VRIEA-PUCV Grant No.  039.428/2017.

\appendix
\section{AdS algebra}
\setcounter{equation}{0}

The algebra of generators $J_{\bA\bB}=-J_{\bB\bA}$ ($\bA,\bB=0,1,\ldots,D$) of AdS group $SO(D-1,2)$ if given by
\bea
[J_{\bA\bB},J_{\bC\bE}]=\eta_{\bB\bC}J_{\bA\bE}+\eta_{\bB\bC}J_{\bA\bE}-\eta_{\bA\bC}J_{\bB\bE}-\eta_{\bB\bE}J_{\bA\bC}\,,\lab{1.1}
\eea
where $\eta_{\bA\bB}=(-1,\underbrace{1,\dots,1}_{D-1},-1)$. Introducing the splitting of indices $\bA=(A,D)$ and with
\bea
P_A&=&J_{AD}\,,\nn\\
J_{AB}&=&-J_{BA}\,,\quad A,B=0,1,\dots,D-1\,,
\eea
the algebra \eq{1.1} (after taking into account that $\eta_{DD}=-1$) takes the familiar form
\bea
&&[P_A,P_B]=J_{AB}\,,\nn\\
&&[P_A,J_{BC}]=\eta_{AB}P_C-\eta_{AC}P_B\,,\nn\\
&&[J_{AB},J_{CE}]=\eta_{BC}J_{AE}+\eta_{AE}J_{BC}-\eta_{AC}J_{BE}-\eta_{BE}J_{AC}\,.
\eea

\section{Variation of LCS action}
\setcounter{equation}{0}

In this appendix we present the non-vanishing parts of the variation of LCS action given by Eq. \eq{4.7},
\be
\d I_{\rm LCS}=\sum_{j=0}^{n}\frac{1}{\r^j}\,\d I_j\,.
\ee
We find the following terms, with $1\leq j\leq (n-2)$:
\bea
\d I_n&=&\ve_{a\ 1\ a_{1}b_{1}\dots a_{n-1}b_{n-1}c}\,\d e^{a}e^{c}K_{-(n-1)}\,,\nn\\
\d I_{n-1}&=& \ve_{ab a_{1}b_{1}\dots d 1 c}\,\d\om^{ab}  e^{c}\nab e^{d}J_{-(n-2)}+\nn\\
&&+\ve_{a1a_{1}b_{1}\dots a_{n-1}b_{n-1}c}\left [\d e^{a}e^{c}K_{-(n-2)}+(\d e^{a}k^{c}- \d k^{a}e^{c})K_{-(n-1)}\right]\,,\nn\\
\d I_j&=&\ve_{1abcd a_{1}b_{1}\dots} \d\om^{ab}\left[ e^{c}\nab e^{d}J_{-(j-1)}- (e^{c}\nab k^{d}- k^{c}\nab e^{d})J_{-j}- k^{c}\nab k^{d}J_{-(j+1)}\right]-\nn\\
&&-\ve_{1ac a_{1}b_{1}\dots a_{n-1}b_{n-1}}\left[\d e^{a}e^{c}K_{-(j-1)}+(\d e^{a}k^{c}- \d k^{a}e^{c})K_{-j}-\d k^{a}k^{c}K_{-(j+1)}\right]\,,\nn\\
\d I_0&=&\ve_{1abcd a_{1}b_{1}\dots}\d\om^{ab}\left[e^{c}\nabla e^{d}J_{1}- (e^{c}\nabla k^{d}- k^{c}\nabla e^{d})J_{0}- k^{c}\nabla k^{d}J_{-1}\right]-\nn\\
&&-\ve_{1ac a_{1}b_{1}\dots a_{n-1}b_{n-1}}\left[\d e^{a}e^{c}K_{1}+(\d e^{a}k^{c}- \d k^{a}e^{c})K_{0}-\d k^{a}k^{c}K_{-1}\right]\,,
\eea
and
\bea
K_{\alpha}&=&\sum\limits_{l=0}^{n-1} \binom{n-1}{l}(R+4ek)^{n-l-1}A_{l\alpha}\,e^{l-\alpha}k^{l+\alpha}\,,\nn\\
J_{\alpha}&=&(n-1)\sum\limits_{l=0}^{n-2}\binom{n-2}{l} (R+4ek)^{n-l-2}A_{l\alpha}\,e^{l-\alpha}k^{l+\alpha}\,,
\eea
where
\bea
A_{l\alpha}=\frac{(-1)^l4^l l!^2}{(2l+1)(l-\alpha)!(l+\alpha)!}\,.
\eea

\section{Alternative proof of the renormalisation theorem}
\setcounter{equation}{0}
In this appendix we show an alternative derivation of the results of  Ref. \cite{19}.
\begin{theorem}
	 A surface counterterm can be added to an action of any classical field theory in the bulk
to cancel all the terms which depend on
	the radial coordinate in an on-shell variation, if any of the following conditions are satisfied:
	\begin{enumerate}
		\item The bulk has the topology $\mathbb{R}\times \partial M$;	
		\item The boundary has a finite number of disjoint pieces and near each one the bulk looks like $\mathbb{R} \times \partial M$.	
	\end{enumerate}
Here, $\partial M$ is any manifold without boundary with the coordinates $x^\a$ and the radial coordinate is labeled by $\r$.  If the fields have asymptotic expansion near the boundary of the form
$\phi^i=\sum_n f_n^i(\rho)\phi_n^i(x^\a),$  where $f_n^i(\rho)$ are functions that depend only on $\rho$ and $\phi_n^i(x^\a)$ are ($\rho$-inde\-pendent) boundary fields,
then the counterterm is a local functional of the boundary fields.
\end{theorem}

Let the action in $D+1$ dimensional bulk $M$ is defined in language of differential forms as
\be\lab{C.1}
S=\int\limits_M L\,.
\ee
A variation of the action \eq{C.1}  takes the form
\be\lab{C.2}
\delta S=\int\limits_M \delta L=\int\limits_M e.o.m. + \int\limits_M d_{D+1}L_D^B
\ee
where \emph{e.o.m} are the terms proportional to the  equation of motion. Formula \eq{C.2} is also valid  without integral and it will be used in that form later.
By using the Stoke's theorem, we can write the last term in \eq{C.2}  as
\be
\int\limits_M d_{D+1}L_D^B=\int\limits_{\partial M} L_D\,,
\ee
where the  boundary of $M$  is placed at fixed distance $\rho=\ve$ near (but not equal) zero and $L_D:=\left.L_D^B\right |_{\r=\ve}$.  Let $\partial M$ be a boundary at each $\rho$.
The most general $D$-form $L_D$ near the boundary is
\be\lab{C.4}
L_D^B=L_D + d\rho \wedge V\,,
\ee
where $V$ is an arbitrary $(D-1)-$form. The exterior derivative in the bulk can be decomposed near the boundary as
\be\lab{C.5}
d_{D+1}=\partial_{\rho} d\rho +d\,,
\ee
where $d$ is the exterior derivative at the boundary and $d_{\rho}$ is the derivative along the direction $\rho$. From Eqs. \eq{C.2}, \eq{C.4} and \eq{C.5}, we get on-shell
\be
\d L = d\rho \wedge \partial_{\rho}L_D -d\rho \wedge dV\,.
\ee
Equivalently,  this can be rewritten as
\be\lab{C.7}
\partial_{\rho}L_D = \delta U +dV\,
\ee
where $\d L=d\r\wedge \d U$. Hence, from \eq{C.7} it follows that
\be
L_D = \delta A +dB+R(x^\a)\,
\ee
where $A=\int d\r U$, $B=\int d\r V$ and $R(x^\a)$ does not depend on $\r$.   This conclusion is valid under  the assumption that the right side of the
Eq. \eq{C.7} is integrable and that the derivative and integral
mutually commute. Therefore, $L_D$ is a sum of a total variation, exact form and a function which does not depend on $\r$.

Consequently, we get
\be\lab{C.9}
\int\limits_{\partial M} L_D=\delta \int\limits_{\partial M} A +\int\limits_{\partial M} R\,,
\ee
where we used the fact that an integral of the exact form $dB$  vanishes due to the Stoke's theorem and because the boundary of a boundary is an empty set. After substituting \eq{C.9}
into \eq{C.2} we obtain on-shell
\be\lab{C.10}
\d (S-S_{\rm ct})= \int\limits_{\partial M} R\,,
\ee
where $S_{\rm ct}=\int_{\partial M} A$.
Since $R$ is  $\r-$independent, the expression  \eq{C.10} is well-defined at the boundary $\r=0$. Thus,
all $\r-$dependent terms can be eliminated by adding  a suitable counterterm. An important observation is that this counterterm is \emph{unique}. Given an  asymptotic solution of the field equations, a near-boundary behavio is fixed.
Furthermore, the counterterm is obtained from th Lagrangian, thus it depends on the same parameters. In other words, we do not include new parameters in the theory. If the starting Lagrangian has
a finite number of parameters, so it does the renormalised Lagrangian.

As the counterterm is obtained as a primitive function of local functions, it is not necessarily  local. The near-boundary expansion method is, however, able to determine only local counterterms.


\end{document}